# Gravitation Field Dynamics In Jeans Theory

A.A. Stupka

*Dnipropetrovsk National University, Quantum Chromoplasma Laboratory,*

*Naukova Str., 13, 49050, Dnipropetrovsk, Ukraine. e-mail:* antonstupka@mail.ru

Abstract. Closed system of time equations for nonrelativistic gravitation field and hydrodynamic medium was obtained taking into account binary correlations of the field, which is the generalization of Jeans theory. Distribution function of the system was build on the base of the Bogolyubov reduced description method. Calculations were carried out up to the first order of a perturbation theory in interaction. Adiabatic and enthropic types of perturbations were corrected. Two new types of perturbation were found.

Jeans theory is the fundamental base for self-gravitating gas studying (Zeldovich & Novikov 1983) and most widespread through the simplicity of the model. Last time generalizations of the Jeans theory for more complicated gas model are actively studied, for example, the model with Burnett terms (Garcia-Colin & Sandoval-Villalbazo 2005), magnetization (Lee & Hong 2007), heat flux (Sandoval-Villalbazo & Garcia-Perciante 2007), binary systems (Tsiklauri 1998). In this paper generalization was built from the other hend, by taking into account freedom degrees of the nonrelativistic longitudinal gravitation field. New independent variables (second correlations of the field) are introduced. This gives a possibility to estimate the influence of the mentioned field on perturbation motion in the gas. This problem is analogous to the Coulomb electronic plasma with Langmuir (plasma) waves (Sokolovsky & Stupka 2006).

## HAMILTON FUNCTION

We will begin with nonrelativistic action for the system of massive particles and Newtonian field (Landau & Lifshitz 1980)

$$S = \int\int \left( \frac{\rho \upsilon^2}{2} - \rho\varphi - \frac{(\nabla\varphi)^2}{8\pi G} \right) dVdt \qquad (1)$$

Let us make the replacement of generalized field coordinate

$$\varphi = \partial_t \lambda . \qquad (2)$$

If we consider equation (1) as nonrelativistic limit of general relativity action, then $\varphi = -c^2 h_{00} / 2$, where $h_{00}$ is small addition term to the proper component of Galilei metrics.

The coordinates shift by small four-vektor $\xi_\alpha$ (the Greek indexes are 0,1,2,3) causes change of metric tenzor (Landau & Lifshitz 1980) $h'_{\alpha\beta} = h_{\alpha\beta} - \frac{\partial \xi_\alpha}{\partial x^\beta} - \frac{\partial \xi_\beta}{\partial x^\alpha}$, from where by comparison with equation (2) one can see $\xi_0 = \frac{\lambda}{c}$. This means that equation (2) corresponds to returning from the coordinate system with $h_{00} = 0$ to the laboratory system.

Let us denote $\boldsymbol{A}=\boldsymbol{\nabla}\lambda$, were the components of $\boldsymbol{A}$ coinside with $-ch_{0i}$ of the system with $h_{00}=0$ at $\xi_i=0$: $-ch_{0i}=\boldsymbol{\nabla}_i\lambda$. In a linear theory the gravitation vector potential is formed from the components $h_{0i}$. Similarly one could name the vector $\boldsymbol{A}$ as longitudinal vector potential, however one should remember that it is a new generalized coordinate of the field exactly in laboratory system.

Then in the second term in equation (1) integrating by parts over time and using equation of continuity

$$\partial_t\rho+\boldsymbol{\nabla}\boldsymbol{j}=0, \tag{3}$$

and after that integrating by parts over space, and in the third term simply changing the order of derivatives we obtain

$$S=\int\int\left(\frac{\rho\boldsymbol{\upsilon}^2}{2}-\boldsymbol{j}\boldsymbol{A}-\frac{(\partial_t\boldsymbol{A})^2}{8\pi G}\right)dVdt \tag{4}$$

In equation (4), unlike equation (1), the generalized coordinate of gravitation field is dynamic one and it is easy to build the Hamiltonian function by standard procedure (Sokolovsky & Stupka 2006). Let us introduce the correspondent velosity $\boldsymbol{E}=\partial_t\boldsymbol{A}$. And nonrelativistic Hamilton function of the system has the form

$$\hat{H}=\hat{H}_0+\hat{H}_{\text{int}}=\left(\hat{H}_{\text{m}}+\hat{H}_{\text{f}}\right)+\left(\hat{V}_1+\hat{V}_2\right),\qquad \hat{H}_{\text{f}}=-\int d^3\boldsymbol{x}\frac{\boldsymbol{E}(\boldsymbol{x})^2}{8\pi},\qquad \hat{V}_1=-\int d^3\boldsymbol{x}\hat{A}_n(\boldsymbol{x})\hat{j}_{on}(\boldsymbol{x}),$$

$$\hat{V}_2=\frac{1}{2}\int d^3\boldsymbol{x}\hat{A}^2(\boldsymbol{x})\hat{\rho}(\boldsymbol{x});\qquad \hat{j}_{on}(\boldsymbol{x})=\sum_a\hat{j}_{oan}(\boldsymbol{x})=\hat{\pi}_{on}(\boldsymbol{x}), \tag{5}$$

where $\hat{H}_{\text{m}}$ and $\hat{H}_{\text{f}}$ are Hamilton functions for free medium particles and longitudinal gravitation field respectively ($\text{rot}\,\boldsymbol{E}=0$). The last expression (5) contain free (i.e. without interaction) momentum density $\hat{\pi}_{on}(\boldsymbol{x})$ (such values have additional subscript $o$), that involve a question about gauge invariance, wich will be solved in standart way (Sokolovsky & Stupka 2006, Landau & Lifshitz 1980).

## DISTRIBUTION FUNCTION

The considered system will be described by field strength $\eta_{1nx} \equiv \boldsymbol{E}_n(\boldsymbol{x},t)$, its longitudinal vector potential $\eta_{2nx} \equiv \boldsymbol{A}_n(\boldsymbol{x},t)$, binary correlations $(\hat{\boldsymbol{E}}_n(\boldsymbol{x})\hat{\boldsymbol{E}}_l(\boldsymbol{x}'))^t$ (we denote them as variables $\eta_\alpha(t)$) and by densities of energy $\zeta_0(\boldsymbol{x},t) \equiv \varepsilon(\boldsymbol{x},t)$, momentum $\zeta_n(\boldsymbol{x},t) \equiv \boldsymbol{\pi}_n(\boldsymbol{x},t)$, mass $\zeta_4(\boldsymbol{x},t) \equiv \rho(\boldsymbol{x},t)$ of gas subsystem (variables $\zeta_\mu(\boldsymbol{x},t)$). Similar parameter set was used in (Sokolovsky & Stupka 2006). Gauge invariant densities of momentum and energy can be introduced in usual manner. For example, gauge invariant mass velocity $u_{an}$ of the electron subsystem is given by the expression

$$\boldsymbol{u}_n(\boldsymbol{x},t) = \boldsymbol{u}_{on}(\boldsymbol{x},t) - \boldsymbol{A}_n(\boldsymbol{x},t)\rho(\boldsymbol{x},t), \tag{6}$$

where $u_{on}$ is free velocity. This allows to use the Galilei transformation and to obtain the results in a gauge invariant form.

Equations of motion for microscopic values of reduced description parameters in terms of the gauge invariant densities have the standart form

$$\dot{\hat{\boldsymbol{E}}}_n(\boldsymbol{x}) = 4\pi G\hat{j}_n(\boldsymbol{x}),\quad \dot{\hat{\boldsymbol{A}}}_n(\boldsymbol{x}) = -\hat{\boldsymbol{E}}_n(\boldsymbol{x});\quad \dot{\hat{\rho}}(\boldsymbol{x}) = -\frac{\partial \hat{j}_l(\boldsymbol{x})}{\partial \boldsymbol{x}_l},\quad \dot{\hat{\varepsilon}}(\boldsymbol{x}) = -\frac{\partial \hat{q}_n(\boldsymbol{x})}{\partial \boldsymbol{x}_n} + \hat{j}_n(\boldsymbol{x})\hat{\boldsymbol{E}}_n(\boldsymbol{x}),$$

$$\dot{\hat{\pi}}_l(\boldsymbol{x}) = -\frac{\partial \hat{t}_{ln}(\boldsymbol{x})}{\partial \boldsymbol{x}_n} + \hat{\rho}(\boldsymbol{x})\hat{\boldsymbol{E}}_l(\boldsymbol{x}). \tag{7}$$

These formulae allow to write the corresponding microscopic equations for products of values $\hat{\boldsymbol{E}}_n(\boldsymbol{x})$. Averaging such equations with nonequilibrium distribution function of the system $\rho(\eta(t),\zeta_o(t))$ we obtain a closed system of equations for parameters describing the system's state (it is convenient to do this by using non-gauge invariant medium variables $\zeta_o$).

To construct distribution function $\rho(\eta(t),\zeta_o(t))$ for this case one can use the Bogolyubov reduced description method (Akhiezer & Peletminsky 1981) starting from the Liouville equation

$$\partial_t \rho(\eta(t),\zeta_o(t)) = \{\hat{H},\rho(\eta(t),\zeta_o(t))\} \equiv \mathbf{L}\rho(\eta(t),\zeta_o(t)). \qquad (8)$$

Field microscopic values and their correlations satisfy the Peletminsky-Yatsenko condition $\mathbf{L}_m\hat{\eta}_\alpha = -i\sum_{\alpha'} c_{\alpha\alpha'}\hat{\eta}_{\alpha'}$. Using a boundary condition of complete correlation weakening, according to (Akhiezer & Peletminsky 1981, Sokolovsky & Stupka 2005) we obtain the following integral equation for distribution function $\rho(\eta,\zeta_o)$

$$\rho(\eta,\zeta_o) = \rho_q(\eta,)w(\zeta_o) + \int_0^{+\infty} d\tau e^{\tau\mathbf{L}_m}\left\{ e^{\tau\mathbf{L}_f}\left(\mathbf{L}_{\text{int}}\rho(\eta,\zeta_o) + 4\pi G\sum_\alpha \frac{\partial\rho(\eta,\zeta_o)}{\partial\eta_\alpha} j_\alpha(\eta,\zeta_o) - \right.\right.$$

$$\left.\left. -\sum_\mu \int d\boldsymbol{x} \frac{\delta\rho(\eta,\zeta_o)}{\delta\zeta_{\mu o}(\boldsymbol{x})} M_\mu(\boldsymbol{x},\eta,\zeta_o)\right)_{\eta\to e^{i\tau c}\eta} - \rho_q(\eta)\mathbf{L}_m w(\zeta_o)\right\}. \qquad (9)$$

Further it is convinient to considerate the equation (9) in the picture of spatial dependence of microscopic values by Baryakhtar-Peletminsky (Akhiezer & Peletminsky 1981) using distribution function $\rho(\boldsymbol{x},\eta,\zeta_o)$ instead of $\rho(\eta,\zeta_o)$. Then one can choose $w(\boldsymbol{x},\zeta_o)$ in the form of local in $\zeta_{o\mu}(\boldsymbol{x})$ quasi-equilibrium distribution function of the medium and $\rho_q(\eta)$ as arbitrary distribution function of the field. Functions $j_\alpha(\eta,\zeta_0)$, $M_\mu(\boldsymbol{x},\eta,\zeta_o)$ are defined by equations of motion for parameters describing the system

$$\dot{\eta}_\alpha(t) = i\sum_{\alpha'} c_{\alpha\alpha'}\eta_{\alpha'}(t) + 4\pi G j_\alpha(\eta(t),\zeta_o(t)), \qquad \dot{\zeta}_{o\mu}(\boldsymbol{x},t) = M_\mu(\boldsymbol{x},\eta(t),\zeta_o(t)). \qquad (10)$$

In the Baryakhtar-Peletminsky picture equation (9) is solvable in a perturbation theory in small interaction ($\hat{V}_1 \sim \lambda^1$, $\hat{V}_2 \sim \lambda^2$, $\lambda << 1$). The weakness of EM interaction allows to build the perturbation theory in small parameter $\lambda = \Omega / \omega_0$, where $\Omega, \omega_0$ are Jeans and collision frequencies starting from estimations $L_s \sim \lambda^0$, $L_{bo} \sim \lambda^0$, $L_1 \sim \lambda^1$, $L_2 \sim \lambda^2$ for contribution in the Liouville operator (Sokolovsky & Stupka 2005).

It was found that in the local rest reference system of the gas subsystem

$$\rho^{0(0)}(\boldsymbol{x},\eta,\zeta_o) = \rho_q(\eta) w(\zeta_o(\boldsymbol{x})), \qquad w(\zeta_o) = \exp\beta\{\Omega(\beta,\mu) - \hat{H}_m + \mu\hat{M}\};$$

$$\rho^{0(1)}(\boldsymbol{x},\eta,\zeta_o) = \int_{-\infty}^{0} d\tau \int d\boldsymbol{x}' \sum_{\alpha} v_{n\alpha}(\boldsymbol{x}',\tau)\Big(\{\rho_q w(\zeta_o(\boldsymbol{x})), \hat{\eta}_\alpha(\hat{j}_{on}(\boldsymbol{x}'-\boldsymbol{x},\tau) + \hat{\beta}(\boldsymbol{x}'-\boldsymbol{x},\tau)\boldsymbol{u}_n(\boldsymbol{x}))\} +$$

$$+\rho_q \hat{\eta}_\alpha \sum_{\mu} \frac{\partial w(\zeta_o(\boldsymbol{x}))}{\partial \zeta_{o\mu}} \mathrm{Sp_m}\, w(\zeta_o(\boldsymbol{x}))\{\hat{\zeta}_\mu(0), (\hat{j}_{on}(\boldsymbol{x}'-\boldsymbol{x}) + \hat{\beta}(\boldsymbol{x}'-\boldsymbol{x})\boldsymbol{u}_n(\boldsymbol{x}))\} +$$

$$+ w(\zeta_o(\boldsymbol{x}))\beta(\boldsymbol{x}')\boldsymbol{u}_n(\boldsymbol{x}') \sum_{\alpha'} \frac{\partial \rho_q}{\partial \eta_{\alpha'}} \varepsilon_{\alpha'\alpha}\Big), \tag{11}$$

where $\hat{A}_n(\boldsymbol{x})$ and $\hat{j}_{an}(\boldsymbol{x})$ in the Dirac picture and matrix $\varepsilon_{\alpha\alpha'}$ were introduced

$$\hat{A}_n(\boldsymbol{x},\tau) \equiv e^{-\tau \mathrm{L_f}} \hat{A}_n(\boldsymbol{x}) = \sum_{\alpha} \boldsymbol{v}_{n\alpha}(\boldsymbol{x},\tau)\hat{\eta}_\alpha, \quad \hat{j}_{on}(\boldsymbol{x},\tau) = e^{-\tau \mathrm{L}_m} \hat{j}_{on}(\boldsymbol{x}); \quad \varepsilon_{\alpha\alpha'} \equiv i \mathrm{Sp_f}\, \rho_q \{\hat{\eta}_\alpha, \hat{\eta}_{\alpha'}\} \tag{12}$$

($\boldsymbol{v}_{n\alpha}(\boldsymbol{x},\tau)$ are known functions), $w$ is equilibrium distribution function of the gas subsystem, $\hat{H}_m$, $\hat{M}$ are the Hamiltonian and mass density of the gas subsystem. Zero superscript in $\rho^{0(0)}(\boldsymbol{x},\eta,\zeta)$, $\rho^{0(1)}(\boldsymbol{x},\eta,\zeta)$ and further indicates, that the corresponding value belongs to the system of the local rest. The first term in equation (11) contains expression

$$\int_{-\infty}^{0} d\tau \int d\boldsymbol{x} \{\hat{A}_n(\boldsymbol{x},\tau)\hat{j}_n(\boldsymbol{x},\tau), \rho_q(\eta) w\} =$$

$$= \int_{-\infty}^{0} d\tau \int d\boldsymbol{x} \Big(\{\hat{A}_n(\boldsymbol{x},\tau), \rho_q(\eta)\}\hat{j}_n(\boldsymbol{x},\tau) w + \rho_q(\eta)\hat{A}_n(\boldsymbol{x},\tau)\{\hat{j}_n(\boldsymbol{x},\tau), w\}\Big), \tag{13}$$

in which the last Poisson brakets can be rewritten in the form [5] (Sokolovsky & Stupka 2005)

$$\{\hat{j}_n(\boldsymbol{x},\tau), w\} = -\beta\{\hat{j}_n(\boldsymbol{x},\tau), \hat{H}_m\} w = \beta \frac{\partial \hat{j}_n(\boldsymbol{x},\tau)}{\partial \tau} w \tag{14}$$

($\beta = T^{-1}$). Then integrating by parts and taking into account the boundary condition of the complete correlation weakening, we see that the lower integral limit disappears and equation (11) takes the form

$$\rho^0(\boldsymbol{x},\eta,\zeta) = \rho_q(\eta)w - \int_{-\infty}^{0} d\tau \int d\boldsymbol{x} \Big( \big\{ \hat{\lambda}_n(\boldsymbol{x},\tau), \rho_q(\eta) \big\} \hat{j}_n(\boldsymbol{x},\tau) w +$$

$$+ w\big(\zeta_o(\boldsymbol{x})\big) \sum_{\alpha} \boldsymbol{v}_{n\alpha}(\boldsymbol{x}',\tau) \beta(\boldsymbol{x}') \boldsymbol{u}_n(\boldsymbol{x}') \sum_{\alpha'} \frac{\partial \rho_q}{\partial \eta_{\alpha'}} \varepsilon_{\alpha'\alpha} \Big) -$$

$$-\beta \int_{-\infty}^{0} d\tau \int d\boldsymbol{x} \rho_q(\eta) \hat{\boldsymbol{E}}_n(\boldsymbol{x},\tau) \hat{j}_n(\boldsymbol{x},\tau) w - \beta \int d\boldsymbol{x} \rho_q(\eta) \hat{\lambda}_n(\boldsymbol{x}) \hat{j}_n(\boldsymbol{x}) w \qquad (15)$$

The last term in equation (15) will be dismissed with term from $\hat{V}_2$.

## EQUATIONS OF MOTION AND PERTURBATIONS

To simplify the further discussion we will keep in equations terms up to the first order in $\lambda$.

$$\frac{\partial \rho}{\partial t} = -\frac{\partial \rho \boldsymbol{u}_n}{\partial \boldsymbol{x}_n}, \qquad \frac{\partial \boldsymbol{u}_l}{\partial t} = -\boldsymbol{u}_n \frac{\partial \boldsymbol{u}_l}{\partial \boldsymbol{x}_n} - \frac{1}{\rho} \frac{\partial (p + p^{(1)})}{\partial \boldsymbol{x}_l} + \boldsymbol{E}_l + \frac{1}{\rho} \big( \beta \hat{\boldsymbol{E}}_l \big)^0,$$

$$\frac{\partial \varepsilon^0}{\partial t} = -\frac{\partial (\varepsilon^0 \boldsymbol{u}_n + \boldsymbol{q}_n^{0(1)})}{\partial \boldsymbol{x}_n} - p \frac{\partial \boldsymbol{u}_l}{\partial \boldsymbol{x}_l} + \rho \boldsymbol{u}_n \boldsymbol{E}_n + (\hat{j}_n \hat{\boldsymbol{E}}_n)^0 . \qquad (16)$$

Here $p$, $p^{(1)}$ are local equilibrium and the first order contribution to pressure of the gas subsystem; $\boldsymbol{q}_n^{0(1)}$ is the first order contribution to energy flux of the gas subsystem. The medium - field correlations are given by the expressions

$$(\hat{j}_n \hat{\boldsymbol{E}}_l) = (\hat{j}_n \hat{\boldsymbol{E}}_l)^0 + (\beta \hat{\boldsymbol{E}}_l) \boldsymbol{u}_n, \qquad \big( \hat{\xi}_{o\mu n} \hat{\boldsymbol{E}}_l \big) = \lambda_{\mu n,\alpha}\big(\boldsymbol{x},\zeta(\boldsymbol{x})\big) (\hat{\eta}_\alpha \hat{\boldsymbol{E}}_l) + S_{\mu n,1l}\big(\boldsymbol{x},\zeta(\boldsymbol{x})\big);$$

$$S_{\mu\, n,\alpha}(\boldsymbol{x},\zeta) = 8\pi \int_{-\infty}^{0} d\tau \int d\boldsymbol{x}' \sum_{\alpha'} \varepsilon_{\alpha\alpha'} \boldsymbol{v}_{l\alpha'}(\boldsymbol{x}',\tau) \mathrm{Sp_m}\, w(\zeta) \big( \hat{j}_{ol}(\boldsymbol{x}',\tau) + \beta(\boldsymbol{x}',\tau) \boldsymbol{u}_l - \rho \boldsymbol{u}_l \big) \hat{\xi}_{o\mu n}(\boldsymbol{x}),$$

$$\lambda_{a\mu\, n,\alpha}\left(\boldsymbol{x},\zeta\right)=-\beta\int_{-\infty}^{0}d\tau\int d\boldsymbol{x}'v_{l\alpha}\left(\boldsymbol{x}'\right)Sp_m w^o\left(\zeta^o\right)\hat{\zeta}_{oa\mu n}\left(\boldsymbol{x}\right)\left(\sum_b \hat{\boldsymbol{j}}_{obl}\left(\boldsymbol{x}',\tau\right)+\hat{\beta}\left(\boldsymbol{x}',\tau\right)u_l\right). \qquad (17)$$

The first order contribution to hydrodynamic fluxes has the form $\zeta^{(1)}_{oa\mu\, n}=\lambda_{a\mu\, n,1l}\left(\boldsymbol{x},\zeta\left(\boldsymbol{x}\right)\right)\boldsymbol{E}_l$, where function $\lambda_{a\mu\, n,\alpha}$ is defined in equation (17). In the considered approximation the relations (11), (14) give the following equations for average gravitation field and correlations of the field

$$\frac{\partial \boldsymbol{E}_n}{\partial t}=4\pi G\rho\,\boldsymbol{u}_n\,, \qquad \frac{\partial(\hat{\boldsymbol{E}}_n\hat{\boldsymbol{E}}_l)}{\partial t}=4\pi G\left(\hat{\boldsymbol{j}}_n\hat{\boldsymbol{E}}_l\right) \qquad (18)$$

From equation (17) it follows that we received a closed system of equations (16), (18) for gravitation field and gas. These equations can be applied for studying field dinamics and correlation phenomena in the system. Supposing the correlations to have thear equilibrium value $(EE)_{nn}=4\pi T$, one obtains equations of Jeans theory, but the first equation (18) allows zero solution, which was forbiden by Poisson equation.

Further we will neglect of spatial dispersion and consider zero correlation radius approximation using notation of a kind $(\hat{\boldsymbol{E}}_n(\boldsymbol{x})\hat{\boldsymbol{E}}_l(\boldsymbol{x}'))=(\hat{\boldsymbol{E}}\hat{\boldsymbol{E}})_{nl}(\boldsymbol{x})\delta\left(\boldsymbol{x}-\boldsymbol{x}'\right)$. Linearization of equations (16) - (18) near the equilibrium after the Fourier transformation gives the following equations for deviations of the corresponding values from the equilibrium

$$\frac{\partial\rho}{\partial t}=-i\rho_0 ku\,,\ \frac{\partial u}{\partial t}=-\frac{i}{\rho_0}k(p_T T+p_\rho\rho)+E\,,\ \varepsilon^0_T\frac{\partial T}{\partial t}=-ik\lambda_4 E-\left(w-\varepsilon^0_\rho\rho_0\right)iku+\lambda_2\left\{(EE)_{nn}-4\pi T\right\}\,; \qquad (19)$$

$$\frac{\partial E}{\partial t}=4\pi G\rho_0 u\,,\ \frac{\partial(EE)_{nn}}{\partial t}=4\pi\lambda_1\left\{(EE)_{nn}-4\pi T\right\} \qquad (20)$$

(we do not use a special notation for the deviations; $\rho_0$ is equilibrium mass density of electrons). Here only longitudinal parts of velosity vector fields were kept, constant coefficients $\lambda_1\delta_{nl}=-G\int d\boldsymbol{x}\lambda_{4n,1l}\left(\boldsymbol{x},\zeta\right)$, $\lambda_2\delta_{nl}=\lambda_{4n,1l}\left(\boldsymbol{x}=0,\zeta\right)$, $\lambda_4\delta_{nl}=\int d\boldsymbol{x}\lambda_{0n,1l}\left(\boldsymbol{x},\zeta\right)$ were introduced and notation $p = p_T T + p_\rho \rho$, $\varepsilon^0 = \varepsilon_T^0 T + \varepsilon_\rho^0 \rho$ was used. So, we have five equations (19) and (20), which form a closed system. It allows to study the influence of average field and field correlations degrees of freedom on such system. And we obtaine two new perturbation types: field one, that characterized by $\lambda_4$ and correlation one, that characterized by $\lambda_1$ and $\lambda_2$. To solve the system we standartly suppose the time dependence to have the form $\sim \exp(\omega t)$. After some calculation we obtain the following corrected dispersion law for the known adiabatic types (Zeldovich & Novikov 1983) for small wave vector and small interaction:

$$\omega = \pm\left(\Omega + O(\lambda^2)\right) + \left(\mp u^2/2\Omega - \Delta/2 + \psi\theta/2\Omega^2 + O(\lambda^2)\right)k^2 + O(k^3) \qquad (21)$$

where $\Omega = \sqrt{4\pi G m}$ is Jeans frequency, $\Delta \equiv 4\pi\lambda_4\, p_T/\varepsilon_T^0$, $\psi \equiv p_T\,(\varepsilon^0 + p - \varepsilon_\rho^0 \rho)/\rho\varepsilon_T^0$, $\theta \equiv 4\pi\lambda_2/\varepsilon_T^0$ and $u = \sqrt{p_\rho + \psi}$ is the sound velocity. For the field type: $\omega = 0$. And for enthropic and correlation types:

$$\omega = \left(\frac{\vartheta - \theta \pm |\vartheta - \theta|}{2} + O(\lambda^2)\right) \mp \left(\frac{\psi\theta\left(\vartheta - \theta \pm |\vartheta - \theta|\right) + \Delta\left(\vartheta + \theta \pm |\vartheta - \theta|\right)\Omega^2}{2\Omega^2\,|\vartheta - \theta|} + O(\lambda^2)\right)k^2 + O(k^3) \qquad (22)$$

where $\vartheta = 4\pi\lambda_1$. Together with the sound velocity the expression (21) contains new terms connected with average field and field correlations because kinetic coefficients $\lambda_4$ and $\lambda_2$, which can be a leading contribution compared with an ordinary expression (Zeldovich & Novikov 1983). Expression (22) gives essentially new effect dew to new kinetic coefficients $\lambda_4$, $\lambda_1$ and $\lambda_2$.

## CONCLUSIONS

We have generalized Jeans theory by taking into account binary correlations of nonrelativistic gravitation field and using time equation for the average field. This have been done by using Bogolyubov reduced description method and mass conservation law starting from action for nonrelativistic gas and Newtonian field. As a result the solution of linearized system have been proposed, that corrected known adiabatic perturbation eigenvalues and introduse two new ones, one of wich is zero, but another is coupled with corrected enthropic perturbation. This, for example, changes the Jeans mass and may be usefull in galaxy organization theories.